# The heterogeneous spatial and temporal patterns of behavior of small pelagic fish in an array of Fish Aggregating Devices (FADs)

M. Capello[1*], M. Soria[1], P. Cotel[1], G. Potin[2], L. Dagorn[3] and P. Fréon[4]

1 UMR EME, Institut de Recherche pour le Développement (IRD), Saint Denis, La Réunion, France
2 ECOMAR, Laboratoire d'Ecologie Marine, Université de La Réunion, Saint Denis, La Réunion, France
3 UMR EME, Institut de Recherche pour le Développement (IRD), Victoria, Seychelles
4 UMR EME, Institut de Recherche pour le Développement (IRD), Sète, France
∗ E-mail: manuela.capello@ird.fr

## Abstract

Identifying spatial and temporal patterns can reveal the driving factors that govern the behavior of fish in their environment. In this study, we characterized the spatial and temporal occupation of 37 acoustically tagged bigeye scads (*Selar Crumenophthalmus*) in an array of shallow Fish Aggregating Devices (FADs) to clarify the mechanism that leads fish to associate with FADs. A comparison of the number of visits and residence times exhibited by the fish at different FADs revealed a strong variability over the array of FADS, with the emergence of a leading FAD, which recorded the majority of visits and retained the fish for a longer period of time. We found diel variability in the residence times, with fish associated at daytime and exploring the array of FADs at nighttime. We demonstrated that this diel temporal pattern was amplified in the leading FAD. We identified a 24-hour periodicity for a subset of individuals aggregated to the leading FAD, thus suggesting that those fish were able to find this FAD after night excursions. The modeling of fish movements based on a Monte Carlo sampling of inter-FAD transitions revealed that the observed spatial heterogeneity in the number of visits could not be explained through simple array-connectivity arguments. Similarly, we demonstrated that the high residence times recorded at the leading FAD were not due to the spatial arrangement of individual fish having different associative characters. We discussed the relationships between these patterns of association with the FADs, the exploration of the FAD array and the possible effects of social interactions and environmental factors.

## Introduction

Pelagic fish species naturally aggregate around floating objects, and fishermen have been constructing and deploying Fish Aggregating Devices (FADs) for decades to attract and catch the fish (Gooding and Magnuson 1967, Fréon and Misund 1999, Dempster and Taquet 2004). Due to the great importance of FADs in tuna fisheries, most of the studies devoted to the relationship between fish and FADs have focused on tuna (Dempster and Taquet 2004). However, tuna are not the only FAD-associated species (Hunter and

Mitchell. 1968, Deudero et al. 1999; Taquet et al. 2007a); several aggregations are multispecific and include the presence of small pelagic fish (Rountree 1990, Fréon and Misund 1999). This multispecificity indicates that the association mechanism could be a universal feature and thus trackable throughout different case studies. The bigeye scad (*Selar Crumenophthalmus*) is one of the main small pelagic fish known to associate with FADs (Taquet et al. 2007a). In recent years, the reported average yearly catch of this species worldwide amounts to 150,000 t (FAO 2010). The catches are usually performed by small-scale fisheries (Potier and Sadhotomo 1995); despite their relatively low catches, these fisheries can be economically important at a local scale, especially for food security.

Previous studies on the behavior of bigeye scads around FADs revealed that FADs increase the encounter rate of bigeye scads and enhance their schooling behavior (Soria et al. 2009), thus validating the "meeting point hypothesis" (*sensus* Dagorn and Fréon 1999 and Fréon & Dagorn 2000). Recent analyses based on high-precision acoustic tagging data which quantifies the dynamics of aggregations of bigeye scads around a FAD (Capello et al 2011) has reinforced the idea that social interactions could play a crucial role in the association between bigeye scads and FADs. At the scale of a whole array of FADs, these interactions could induce the concentration of the associated biomass in a subregion of the array. However, no study has yet investigated if the fish, at least for a given period and for a given array, select a particular (or a subset of) FAD(s) or if they rather occupy a FAD array homogeneously. In this paper, we analyzed the behavior of 37 acoustically tagged bigeye scads in an array of shallow coastal FADs, each equipped with acoustic receivers (listening stations). We assessed the amount of time that the fish spent around each FAD and characterized their movements over the whole array to gain knowledge of the association process, with a particular emphasis paid to the emergence of spatial and temporal patterns.

**Material and methods**

**Study Site and Stationary acoustic array**

The study was conducted at Saint Paul's Bay, which is located on the northwest coast of La Réunion Island, South Western Indian Ocean. Fifteen different artificial structures (anchored buoys, fishing anchored FADs or wrecks) were present in the bay and anchored at 15 to 50 m depth. A detailed description of the designs of these structures and their location within the bay can be found in Soria et al. (2009) and in Fig. 1.

To monitor the residence times and the movements of fish in this array, we used coded acoustic tags and receivers (VR2s) (Vemco, Halifax, Canada). We instrumented each structure or group of packed structures in the bay with an acoustic receiver. Based upon an in-situ range test, we determined that the distance of 150 m was the maximum range of reliability for the acoustic receivers (details in Soria et al. 2009). The distances between some of the artificial structures, particularly those in the central area, were sometimes very short and less than the maximum detection range of a Vemco V7 tag by a VR2 (150 m). Consequently, to limit the overlapping of the VR2's detection areas, we decided to deploy one regrouped receiver near the structures (Fig. 1, VR2/FAD G and H; details in Soria et al. 2009). A total of 9 VR2 were deployed in the bay, covering all of the artificial structures. For clarity, these 9 instrumented structures will hereafter be referred to as FADs. The distance between the adjacent instrumented FADs ranged from 310 to 1140 m.

**Experimental protocol**

In May 2006, 105 bigeye scads were captured around the sea fish farm using hand lines. A total of 87 fish were then conveyed to the Aquarium of La Réunion Island and housed in three holding tanks; 18 fish were not large enough to be tagged and were immediately released. After 11 days of acclimation, 46 fish (mean fish size ± SD = 18 ± 1.2 cm (fork length) and mean weight = 102 ± 22 g) were anesthetized and tagged with coded Vemco V7 transmitters (V7-2L-R04K, 69 kHz, random rate of transmission every 70-140 s, rated battery life 130-150 days, length 18.5 mm, diameter 7 mm, mass in water 0.75 g). Details of the fishing, holding conditions, surgical operation and release protocol were detailed in Soria et al. 2009.

Because the bigeye scad is an obligate schooling fish species (Soria et al. 2007), we released the 46 tagged fish with 34 non-tagged conspecific fish so they could immediately form small schools. The fish were transported to the site and, on June 8, 2006, they were gently released in roughly equal numbers at 3 different points (Fig. 1). The acoustic receiver array was operational during the entire 4-month experimental period. Among the 46 tagged fish, 9 were removed from the analysis as they were never detected or only detected at FADs during the first hours after release. We finally worked on a total of 37 tagged fish. Detections were recorded by all of the VR2s in the bay. More than half of the remaining fish (58%) were no longer detected by the VR2 array after day 6 of the experiment. No fish were detected after July 17, 2006, thus setting an upper bound of 40 days for the fish Retention Time in the Array (RTA), i.e., the time elapsed between the first and last detection of fish in the FAD array.

**Data analyses**

Following the method used in previous acoustic telemetry studies on the behavior of fish at FADs (Ohta and Kakuma 2005, Dagorn et al. 2007, Taquet et al. 2007b), we calculated the Continuous Residence Time (CRT) as the time during which a tagged fish was detected by an acoustic receiver without a gap between two acoustic detections superior to a defined Maximum Blanking Period (MBP). Therefore, the MBP corresponds to the maximum time during which a tagged fish is not detected by the receiver but still considered to be around the FAD. The choice for a value of the MBP depends on the tag specifications, and knowledge of the behavior of the fish, including the time scales. In the case of tuna, (Ohta and Kakuma 2005, Dagorn et al. 2007) previous studies adopted a MBP of 24 hours to observe the dynamics of tuna in large-scale arrays of FADs. In the present study, we chose a smaller value for the MBP (MBP=60 minutes) to adapt to the time scales involved in the behavior of the bigeye scad in the array of FADs in St. Paul's bay and to address the fine-scale behavior of the fish. Similarly, we defined a minimum residence time (MRT) that sets a lower bound for the residence times. This time corresponds to the path duration (estimated through a linear movement at a mean speed of 1 body length per second) that a fish needs, in average, to cross the region of detection of a VR2 without stopping. For the bigeye scad, we chose a MRT of 15 minutes. The MRT enabled us to avoid the spurious effects of network anisotropy affecting residence times, which implies a higher probability of cutting residence times in FAD-dense regions. Using these parameters, we pooled the acoustic detections to extract behavioral information.

We then analyzed the fish behavior in our array using two main quantities: survival curves for residence times (Haccou and Meelis 1994) and the number of visits per FAD. In the first part of the analysis (Fig. 2), the number of visits per FAD was only calculated when the fish moved between two different FADs. This allowed us to compare the experimental data with the Monte Carlo simulations (see below), thus taking into account the transition probabilities reflecting the connectivity of the array. In the rest of the paper, we considered all of the possible visits, taking into account both the subsequent visits of fish at the same FAD and those resulting from the transitions among different FADs. A survival curve analysis of the residence times provided information on the probabilities of fish leaving a FAD (or a set of FADs) after a certain amount of time. A comparison among the survival curves was performed using the logrank test with the library *survival* of the **R** statistical programming language (R 2011)

Finally, we characterized individual-fish behavior through the temporal ping pair-correlation among acoustic detections p(t), which is defined as:

$$p(t) = \frac{\sum_{i>j} n(t_i)n(t_j)}{N} \delta(t - |t_i - t_j|)$$

where $n(t_i)$ is equal to 1 (0) when an acoustic detection is (not) present at time $t_i$, and the double sum runs over all of the different pairs of detections. The delta function selects pairs of detections at a relative temporal distance in the interval [$t$, $t+\Delta t$], with $\Delta t$=20 minutes and $N$ being a normalization constant corresponding to the total number of pairs. This quantity was useful for detecting the possible emergence of temporal patterns in the behavior of individual fish.

**Monte Carlo simulations**

Monte Carlo runs were used to simulate the movement of 37 fish during 40 simulated days. The initial conditions were set following the experimental protocol, and the initial positions of the fish were set near the FAD closest to the experimental positions of release. On each day, we verified if the fish were staying or leaving the network according to an exponential probability of fish survival in the array, which was obtained from the fit of the experimental survival curve for the fish RTA (Fig. S1A in Supplementary Information). We then assigned to the "survived fish" a total number of daily visits sampled from the experimental distribution of the number of visits per day (Fig. S1B in Supplementary Information). We simulated the fish dynamics using network-dependent transition probabilities. According to the connectivity of the FAD array, we then assigned different probabilities to inter-FAD movements (see Fig. S2).

Our network was comprised of a southern region with two FADs (A and B), a central FAD-dense region (C-H) and a northern region with one FAD (I). To take this spatial anisotropy into account, we assigned two types of movements to the nearest neighboring FADs: type-0 transitions corresponded to movements between neighboring FADs located within the same region of the array (southern, central and north); type-1 transitions were assigned to movements between remote neighboring FADs, i.e., movements between the neighboring southern and the northern FADs to the central region. Type-2 (Type-n) transitions corresponded to movements between non-neighboring FADs, having 1 (n-1) FAD(s) between them.

The transition probabilities for each type of movement were calculated by dividing the experimental frequency of occurrence of each category of movements by the total number of bonds of each type (see Table ST1).

**Results**

**The Number of visits at different FADs and the comparison with Monte Carlo simulations**

We calculated the number of visits involving transitions between two different FADs for all of the fish during the full period of observation and found a strong spatial variability, with one FAD (FAD H, Fig. 1) concentrating the majority of the visits (Fig. 2). To test if the finite-size statistics and array-connectivity could affect our results, we compared our experimental data with the Monte Carlo simulations of fish displacement in the array.

For most of the FADs (A-E and I), the experimental values of the number of visits matched the simulated values within two standard deviations. However, the model was unable to reproduce the large discrepancies in the number of visits found in the central part of the network, which were nearly doubled around FAD H and highly depleted around FADs F and G.

**Residence times per FAD**

Analyzing the CRTs for all FADs and fish as a function of time, we found a strong temporal and spatial variability in the stay durations at the FADs (Fig. 3). The period between day 1 and day 27 indicated a dominance in the residence times associated with

FAD H. Nevertheless, during this period a lack of visits of unknown origin emerged between day 12 and 21. Therefore, to investigate the origin of the observed spatial heterogeneity, we concentrated on the initial period between days 1 and 11 for the rest of the analysis. This temporal window was characterized by a stable dominance of FAD H and concerned a large amount of fish.

We compared the amount of time spent by the fish at the different FADs, taking survival curves of the CRTs from days 1 to 11. The results provided clear evidence that FAD H was the dominant FAD. Indeed, the survival curves of the fish that stayed associated with FAD H showed longer residence times (Fig. 4). The abrupt drop at approximately 12 hours was due to diel rhythms, as shown below. FAD D showed a mixed survival curve, with few long residence times recorded during the first days of the experiment. The other FADs demonstrated shorter residence times and had similar survival curves (p-value = 0.059 for logrank test between FAD A, B, C, E, I)).

**Day vs. Night behavior**

To further characterize the dynamics of association with the FADs during the selected period of observation, we calculated the number of visits and the survival curves of the CRTs during daytime and nighttime between days 1 and 11. The visits were more uniformly distributed over the full array at nighttime than during daytime, and the total number of visits was more than 50% higher during the nighttime (143 visits during the nighttime versus 91 visits during daytime), suggesting that the nighttime was the period when the fish primarily explored the array (Fig. 5A). The survival curves during the daytime and nighttime (Fig 5B) were significantly different (p-value=0 with logrank test). Much longer stays were recorded during the day than during the night, which is consistent with the previous result. The survival curve of residence times during the daytime was significantly different when the CRTs at FAD H were excluded (p-value= 0.0002 with logrank test between survival curves at daytime with and without FAD H). The difference between the survival curves during the daytime and nighttime was no more statistically significant at the 0.05 level than when the daytime stays at the FAD H where not taken into account (p-value= 0.084 with logrank test between survival curves at daytime and nighttime without FAD H).

**Individual fish behavior**

To see if these CRT patterns were common to all of the fish, we compared the time spent at FAD H relative to the total time spent at all the FADs among individuals. To avoid any bias due to fish leaving the network prematurely, we compared the fish that stayed in the network for at least 11 days.

Indeed, there was an individual variability in the associative behavior of fish (Fig. 6). Three classes were established: "(non-) H-associative" fish for whose residence times at FAD H constituted the "(minority-) majority" of the total association time, and the "H-mixed" fish that spent approximately half of their association time at FAD H. The class non-H-associative fish regroups five individuals that resided primarily in the southwest of the network and associated with FADs A and B.

We then compared the survival curves of the residence times for these three classes of fish (see Fig. 7). We performed the comparison of survival curves with and without considering FAD H. There was obviously a difference in survival curves among the three categories of fish (p-value= 0.0004 with logrank test), with the H-associative fish having longer residence times. However, we found that without the contribution of FAD H, all classes showed the same survival curves (p-value= 0.284 with logrank test).

Finally, we took a deeper look at individual behavior using the temporal pair-correlation among acoustic detections $p(t)$, which was calculated for FAD H and the other FADs.

Among the H-associative fish, four individuals showed a clear 24-hour periodicity when visiting the FADs. Remarkably, it was again FAD H that exhibited special properties because the periodicity was particularly marked for fish staying and visiting this FAD. The fish visited FADs other than H, but with very short stays that were characterized by narrow peaks in $p(t)$ (Fig. 8).

In Fig. 9, we quantified the exploratory behavior of the fish by taking the total number of different FADs visited by a given fish as a function of their association rate. Indeed, there was an increased tendency towards exploration for fish spending longer time at the FADs.

# Discussion

Underwater observations have indicated that bigeye scads, when associated with a FAD, usually stay within a distance of approximately 50-100 m around it, a distance lower than the maximum range of detection of the VR2 (150 m). This was further confirmed by accurate HTI measurements (Capello et al 2011). We were therefore confident in the capability of the VR2s to detect the association between a tagged fish and a FAD. The time scale resulting from the low value retained for MBP (60 min) allowed for a better understanding of the fine-scale dynamics of small-pelagic fish associated with FADs in the FAD array.

In our approach, spatial patterns emerged in the number of visits and the residence times of the fish, with a leading FAD concentrating most of the visits and long residence times during the 11-day study period.

Demonstrating the origin of spatial patterns through a field-based tagging experiment is difficult for many reasons. First, undersampling by itself could produce a non-homogeneous distribution in the presence of a homogeneous system. This issue was crucial in the presence of the small number of tagged fish that showed exponentially decaying RTAs. However, the array connectivity could also induce heterogeneity in the number of visits per FAD. Indeed, the central region of the array concentrated a high density of FADs when compared to the southern and northern regions, where few FADs were present. To evaluate these issues, we ran Monte Carlo simulations that took into account the array connectivity and the experimental finite-size statistics. The fact that our experimental data were several standard deviations different from our model outputs demonstrated that the observed FAD variability was not the result of a simple consequence of finite-sample size, nor was it the result of simple considerations of the array connectivity.

Other possible explanations for the presence of spatial patterns were investigated through an analysis of fish behavior in the period corresponding to the first 11 days, where the largest number of fish was observed and the spatial patterns were stable. It appeared that the fish primarily explored the FAD array during the nighttime (visits were more numerous and more evenly distributed in space than during the daytime) whereas during the daytime, the fish tended to stay associated with the FADs. However, this diel pattern was not the only cause of residence time variability. Remarkably, FAD-H (the leading FAD) amplified these diel patterns and contributed to the majority of long stays during the daytime. To clarify the possible role of individual fish behavior, we then identified three main classes of fish according to the amount of time spent at FAD-H. We showed that the emergence of a leading FAD was not due to the spatial arrangement of individual fish having different tendencies towards association. Indeed, when removing the residence times recorded at FAD H, all of the fish demonstrated the same survival curves. These findings suggest that the observed spatial pattern was more the result of FAD-H (or its surrounding habitat and associated biomass, as discussed below) affecting fish behavior rather than the outcome of a class of individuals with higher associative character located around FAD H.

Three potential reasons could be proposed to explain the emergence of a leading FAD: the intrinsic characteristics of the FAD, the particular micro-habitat surrounding the FAD or social interactions. First, our FADs were not identical in structure, which precludes us from rejecting the first interpretation. Although FADs were more or less all anchored at similar depths, they were composed by different materials and had different shapes. However, we noted that the leading FAD changed from FAD H to FAD D during the course of the experiment (Fig. 3). This finding could support an argument against the presence of spatial patterns due purely to the intrinsic characteristics of FAD H. These results should be confirmed by further experiments because the transition from leading FAD H to FAD D was observed at the end of the experiment, when few fish were present. Indeed, because our overall conclusions are primarily based on small number of tagged individuals, they would deserve further investigation.

Secondly, the emergence of a leading FAD could also arise from a particular micro-habitat around the FAD. It is likely that surrounding conditions such as temperature, current (speed and/or direction), turbidity, prey abundance and other factors were not identical around all of the FADs. Unfortunately, it was not possible to observe the local conditions during the experiment. This clearly emphasizes the need to monitor the physical and biological environment (including the abundance of conspecifics) around the FADs during tagging experiments. Automated equipment should be developed as a full monitoring tool, but it would be difficult to maintain over several days or weeks.

Finally, we observed that the ability of a fish to explore the FAD array increased with its association to FADs, which can be interpreted as evidence of the influence of social interactions on FAD election. Among the fish that were primarily associated with FAD H, some individuals additionally showed daily periodicity in their stays, with regular night excursions. This result could simply reflect different levels of fidelity to the array. However, the relationship between association and exploration could be important for understanding the full mechanisms leading fish to associate with FADs. One possible interpretation is that the social interactions at the leading FAD during the daytime could increase the efficiency of the searching behavior when the fish leave the FAD. This social enhancement effect (known as the "many-wrong principle"; Bergman and Donner 1964; Grunbaum 1998; Codling et al. 2007) could significantly improve the ability of an individual to reach a target point, and it might be particularly important for small pelagic fish, where individuals alone have limited navigational ability. Therefore, characterizing the periodic behavior around FADs, as well as the range of the night excursions of the tagged fish, could provide information about the group sizes associated with the different FADs and, more generally, on the associated biomass within FAD arrays.

The current state of research on FADs requires highly instrumented experimental campaigns and a deeper understanding of individual behavior at a fine scale. Combining several acoustic methods, as well as controlling environmental parameters, is difficult in offshore experiments on large arrays of FADs targeting tuna. Instead, conducting field experiments in shallow water and on small FAD arrays is accessible. Therefore, studying the association mechanisms of small pelagic fish represents a valuable way to investigate the different association hypotheses at a fine scale. We would like to promote this type of research, as we believe that it would provide key information to better understand the interactions between fish and FADs and their effects on fish behavior.

**Acknowledgments**

We are grateful to the team of the Aquarium of Reunion Island. E. Tessier, P. Chabanet, P.

Berthier and L. Berthier are thanked for their assistance in catching and releasing fish and in deploying the acoustic system. G. Lemartin is thanked for his assistance during data acquisition and data processing. This study was funded by the Social European Fund (UE-IFOP), the Regional Council of Reunion Island and the 'Run Sea Science' program (Theme - Capacity Building, contract 229968).

## References

Bergman, G. and Donner, K. O., 1964. An analysis of the spring migration of the common scoter and the long-tailed duck in southern Finland. Acta Zool. Fenn., 105, 1-59.

Capello, M., Soria, M., Cotel, P., Deneubourg, J-L. and Dagorn, L., 2011. Quantifying the Interplay between Environmental and Social Effects on Aggregated-Fish Dynamics. PLoS ONE 6(12): e28109.

Codling, E. A., Pitchford, J. W. and Simpson, S. D., 2007. Group navigation and the "many-wrongs principle" in models of animal movement. Ecology, 88, 1864-1870.

Grunbaum, D., 1998. Schooling as a strategy for taxis in a noisy environment. Evol. Ecol., 12, 503-522.

Dagorn, L. and Fréon, P., 1999. Tropical tuna associated with floating objects: a simulation study of the meeting point hypothesis. Can. J. Fish. Aquat. Sci., 56, 984-993.

Dagorn, L., Holland, K.N. and Itano, D.G., 2007. Behavior of yellowfin (Thunnus albacares) and bigeye (T. obesus) tuna in a network of fish aggregating devices (FAD). Mar. Biol., 151, 595-606.

Dempster, T. and Taquet, M., 2004. Fish aggregation device (FAD) research: gaps in current knowledge and future directions for ecological studies. Rev. Fish Biol. Fish., 14, 21-42.

Deudero, S., Merella, P., Morales-Nin, B., Massuti, E. and Alemany, F., 1999. Fish communities associated with FADs. Sci. Mar., 63, 199-207.

FAO Yearbook: Fishery and Aquaculture Statistics 2008, Rome: FAO, 2010, 72pp.

Fréon, P. and Dagorn, L., 2000. Review of fish associative behaviour: toward a generalisation of the meeting point hypothesis. Rev. Fish Biol. Fish., 10, 183-207.

Fréon, P. and Misund, O.A., 1999. Dynamics of Pelagic Fish Distribution and Behaviour: Effects on Fisheries and Stock Assessment. Fishing News Books (Blackwell Science, London).

Gooding, R. M. and Magnusson, J. J., 1967. Ecological significance of a drifting object to pelagic fishes. Pacific Science, 21, 486-497.

Haccou, P. and Meelis, E., 1994. Statistical analysis of behavioural data. Oxford University Press, 400pp.

Hunter, J. R. and Mitchell, C. T., 1968. Field experiments on the attraction of pelagic fish to floating objects. ICES J. Mar. Sci., 31, 427-434.

Ohta, I. and Kakuma, S., 2005. Periodic behavior and residence time of yellowfin and bigeye tuna associated with fish aggregating devices around Okinawa Islands, as identified with automated listening stations. Mar. Biol., 146, 581-594.

Potier, M. and Sadhotomo, B., 1995. Seiners fisheries in Indonesia. In: M. Potier and S. Nurhakim BIODYNEX: Biology, Dynamics, Exploitation of the Small Pelagic Fishes in the Java Sea, Pelfish, Jakarta, pp. 49–66.

R Development Core Team, 2011. R: A language and environment for statistical computing. R Foundation for Statistical Computing, Vienna, Austria. ISBN 3-900051-07-0, http://www.R-project.org.

Rountree, R. A., 1990. Community structure of fishes attracted to shallow water fish aggregation devices off South Carolina, USA. Environ. Biol. Fishes, 29, 241-262.

Soria, M., Fréon, P. and Chabanet, P., 2007. Schooling properties of an obligate and a facultative fish species. J. Fish Biol., 71, 1257-1269.

Soria, M., Dagorn, L., Potin, G. and Fréon, P., 2009. First field-based experiment supporting the meeting point hypothesis for schooling in pelagic fish. Anim. Behav., 78, 1441-1446.

Taquet, M., Sancho, G., Dagorn, L., Gaertner, J-C., Itano, D., Aumeeruddy, R., Wendling, B. and Peignon, C., 2007a. Characterizing fish communities associated with drifting fish aggregating devices (FADs) in the Western Indian Ocean using underwater visual surveys. Aquat. Living Resour., 20, 331-341.

Taquet, M., Dagorn, L., Gaertner, J-C., Girard, C., Aumerruddy, R., Sancho, G. and Itano, D., 2007b. Behavior of dolphinfish (Coryphaena hippurus) around drifting FAD as observed from automated acoustic receivers. Aquat. Living Resour., 20, 323-330.

**Figures:**

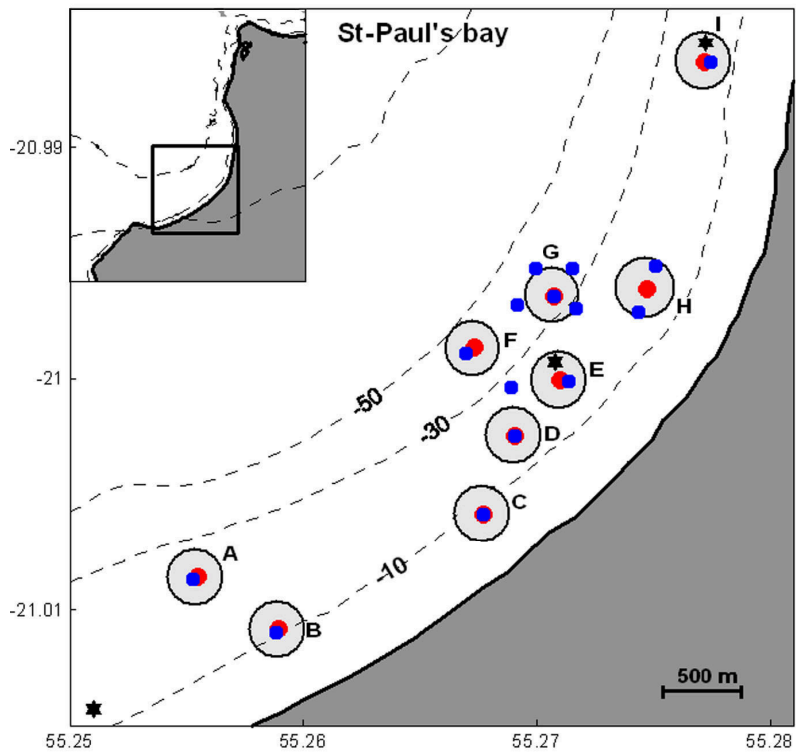

**Figure 1**: Map of Saint Paul's Bay with location of FADs (blue dots). The nine FADs equipped with VR2 Vemco receivers are identified (red disks) with their area of detection (outlined circles). The positions of the fish release points are also shown (∗). The broken line represents isobaths in meters (modified from Soria et al. 2009).

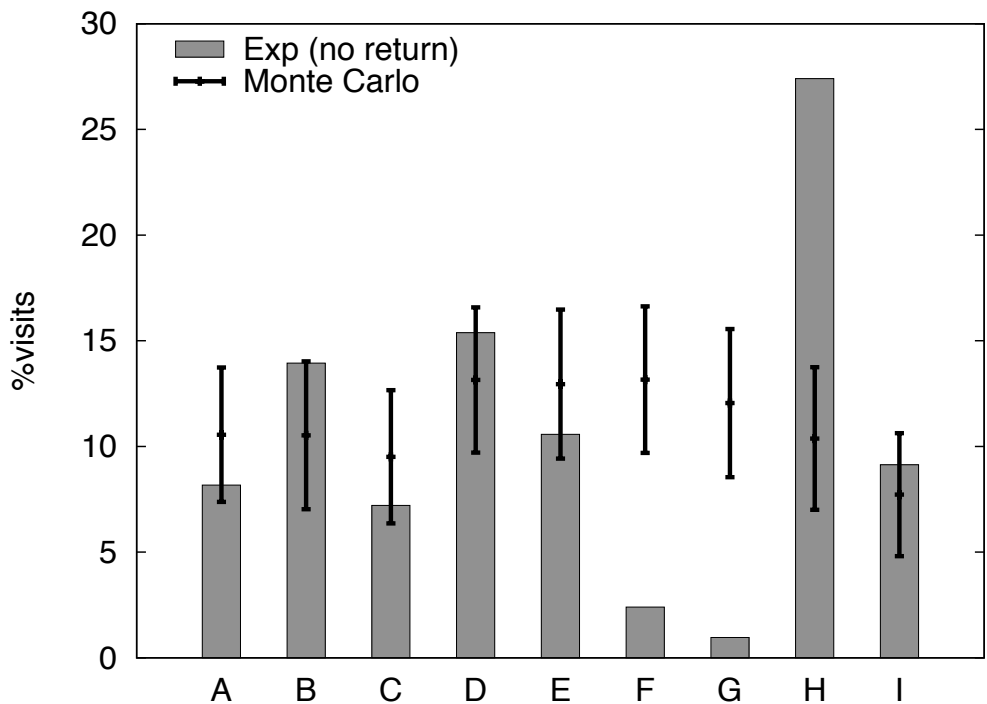

**Figure 2**: Comparison between the experiment (grey histogram) and the Monte Carlo simulations (black points correspond to mean value and bars indicate two standard deviations) to determine the percentage of visits per FAD, which involved transitions among the different FADs (N=208). The model was run 1000 times for 37 fish.

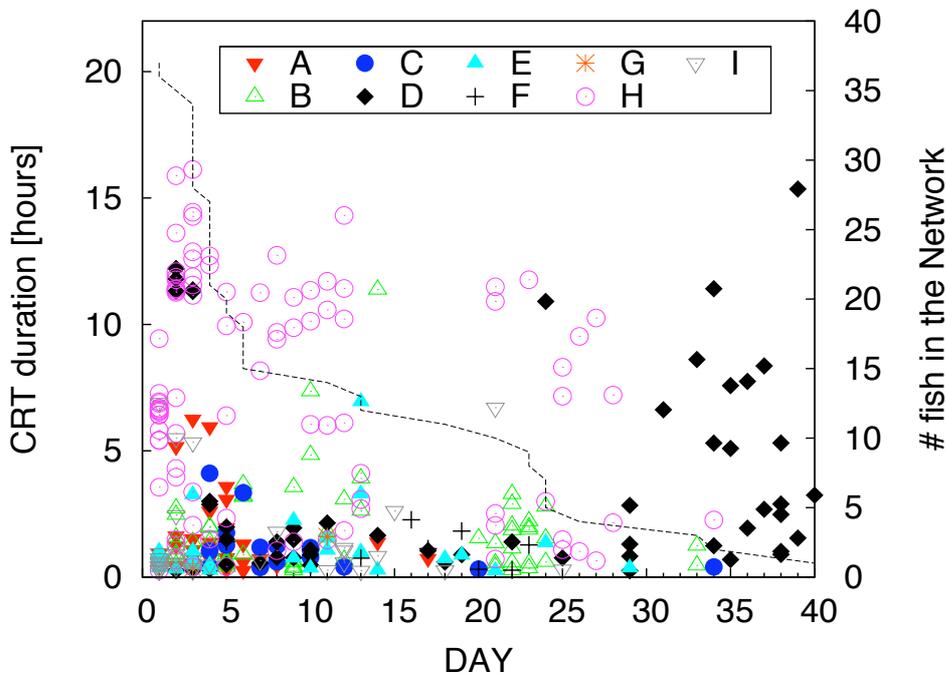

**Figure 3**: The CRT duration (left axis and symbols) at the different FADs and the number of fish present in the network (right axis and dashed line) as a function of time (in days).

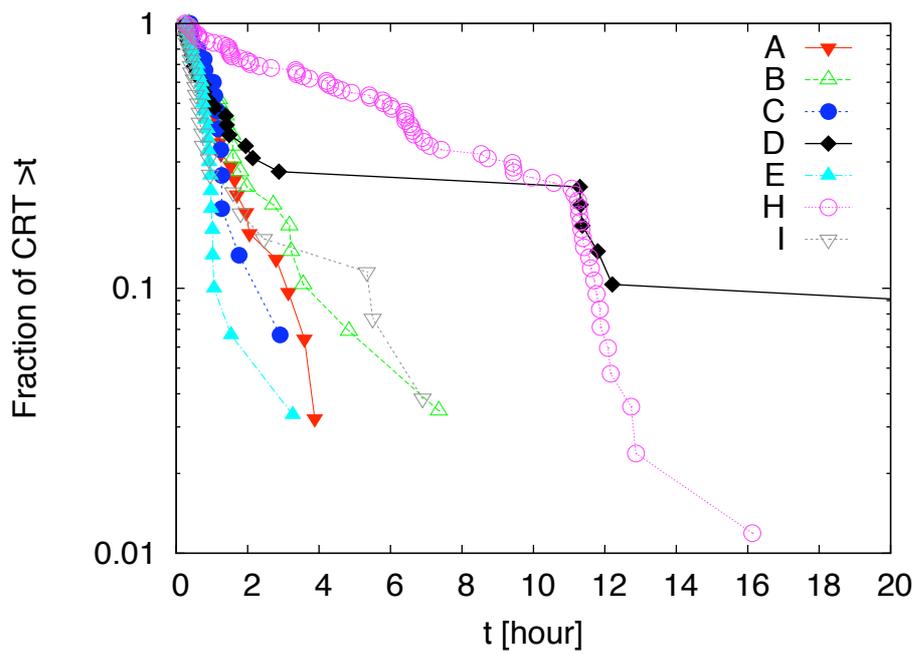

**Figure 4**: The survival curves of the CRT at different FADs during the first 11 days of the experiment in a semi-logarithmic scale.

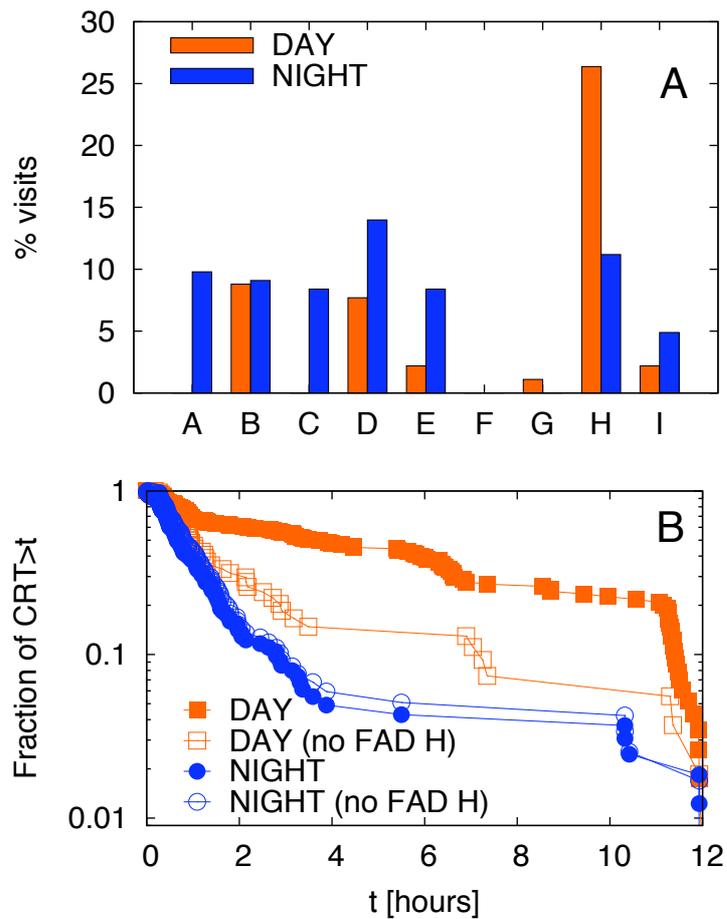

**Figure 5**: The percentage of visits per FAD (A) and the survival curves of the CRTs (B) during daytime (6:00-18:00, red squares) and nighttime (18:00-6:00, blue circles) of all fish and FADs between days 1 and 11. The total number of visits is 91 during the daytime and 143 during the nighttime. In (B), the empty squares and circles indicate the survival curves of the CRTs without FAD H during the daytime and nighttime.

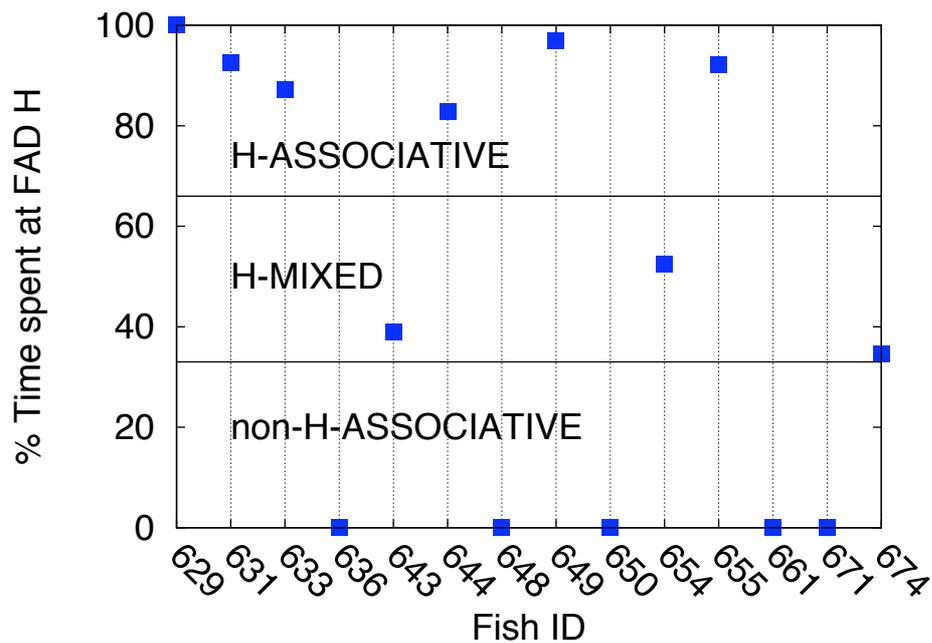

**Figure 6**: The percentage of time spent at FAD H for the different fish. The three categories H-ASSOCIATIVE, H-MIXED and NON-H- ASSOCIATIVE are explained in the text.

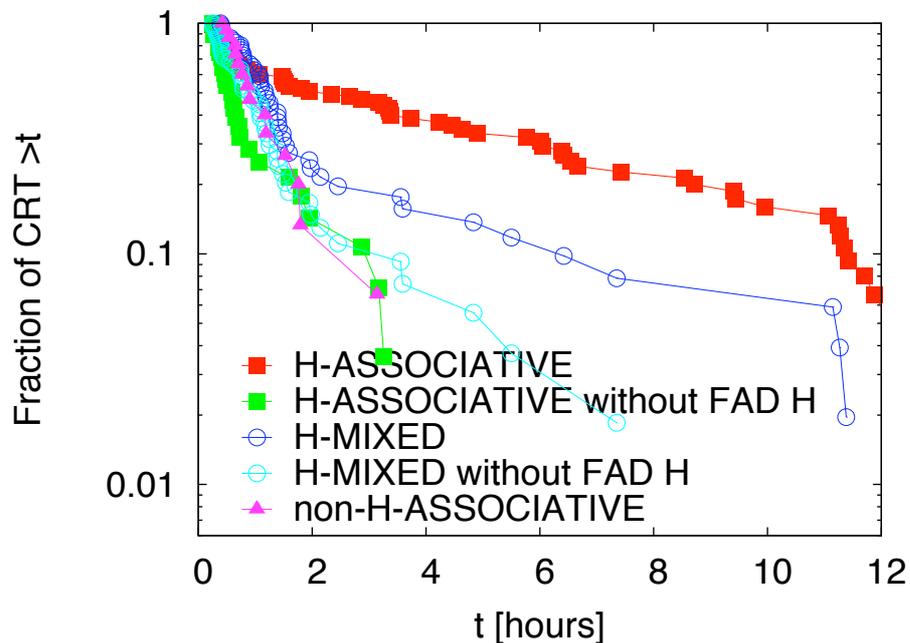

**Figure 7**: The comparison among the survival curves of the CRTs for the three classes of fish: H-associative, mixed and non-H-associative, with and without the CRTs recorded at FAD H.

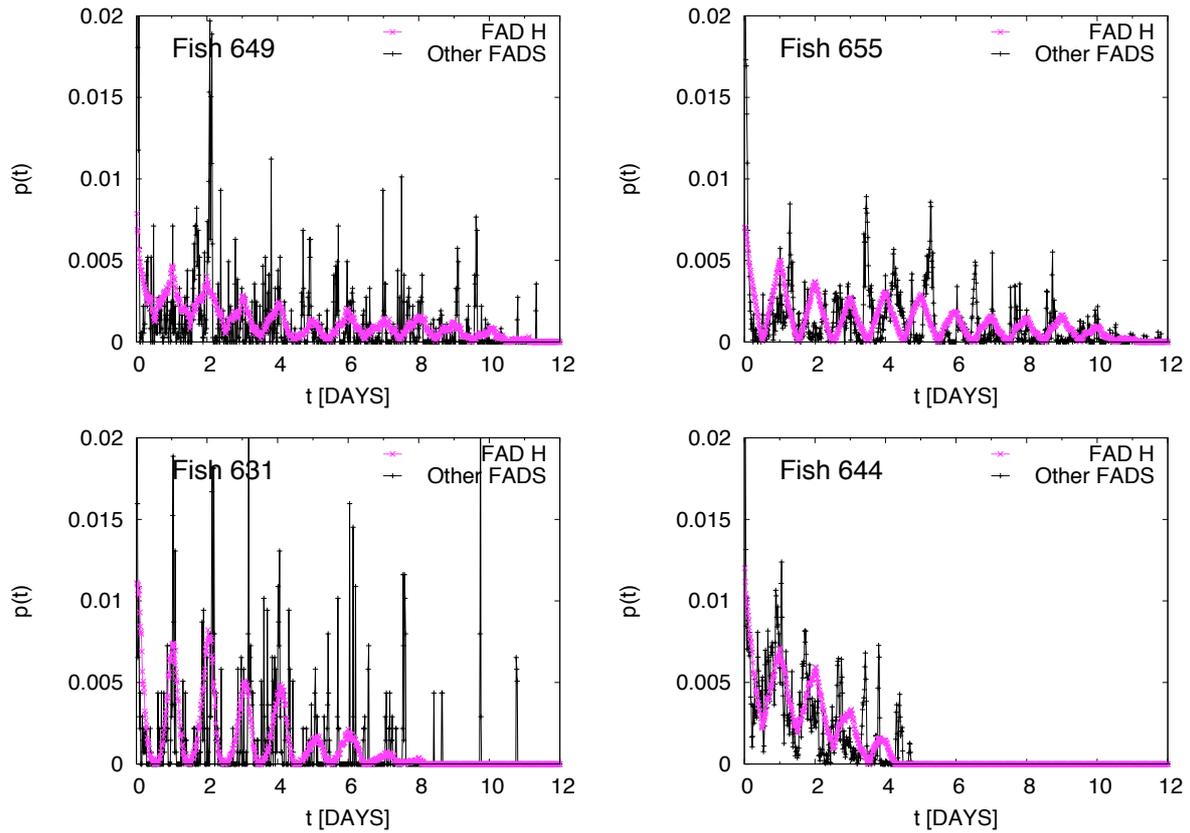

**Figure 8**: The temporal self-correlation function among acoustic detections for four tagged fish residents of the dominant FAD H.

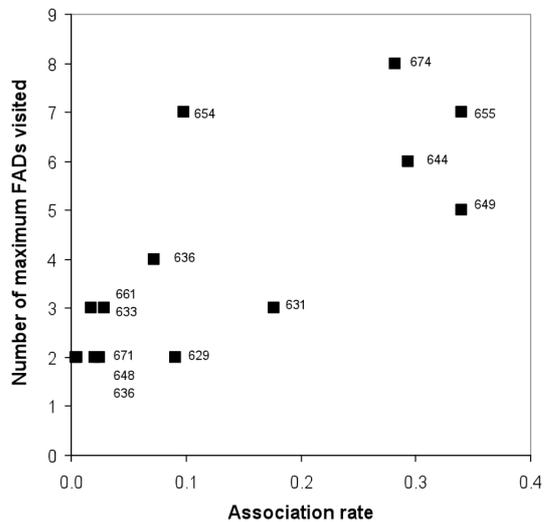

**Figure 9**: A plot of the number of different FADs visited by individual tagged fish (each point corresponds to an ID number) versus their association rate between days 1 and 11.

**Supplementary Figures and Supplementary Tables**

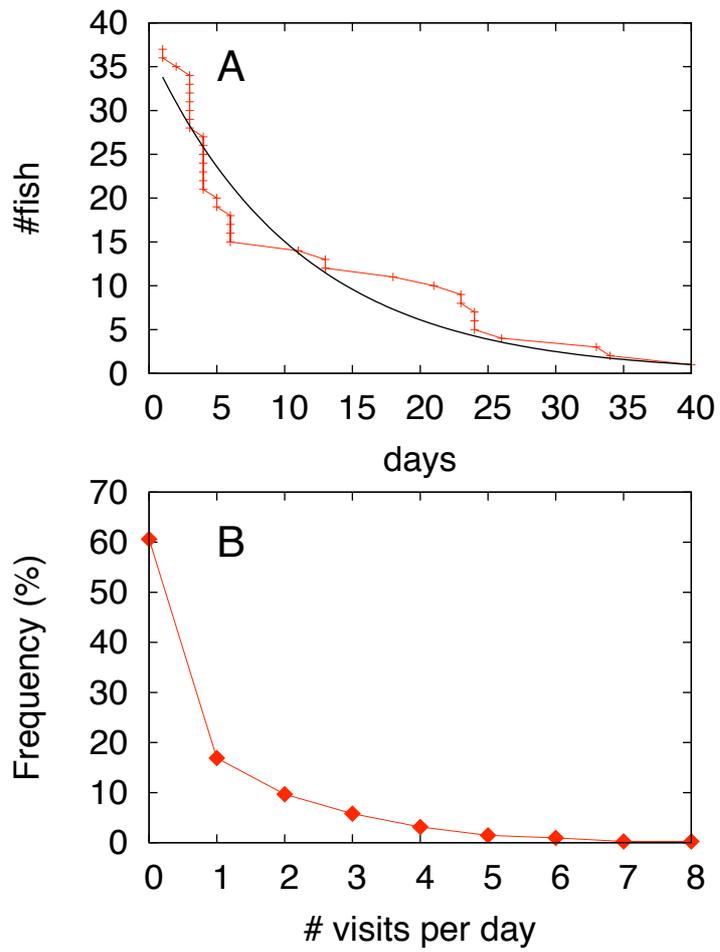

**Figure S1**: A) The survival curve of fish RTA in the FAD network and fit with f(x)=37*exp(-0.09*x). B) The distribution of the number of visits per fish per day.

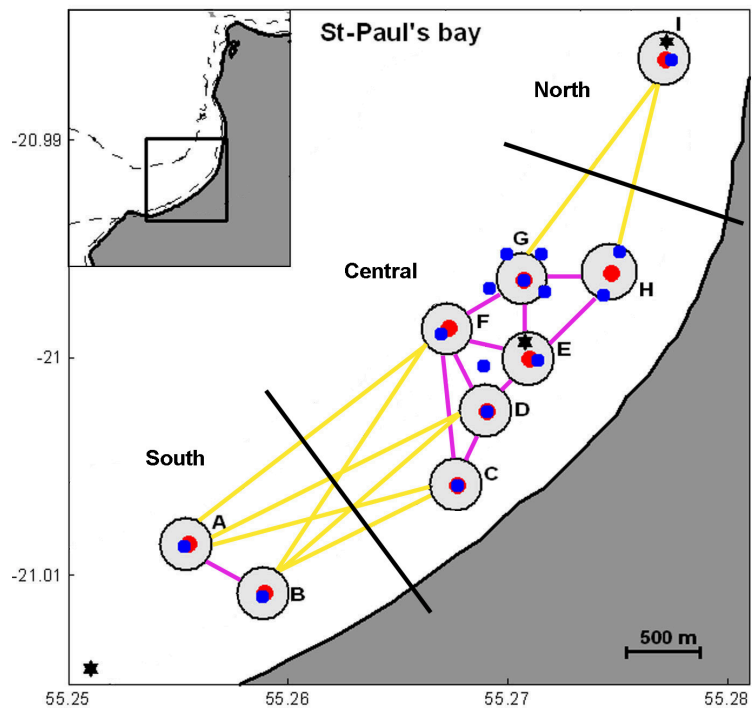

**Figure S2**: The map of Saint-Paul's Bay as in Fig. 1 with Type 0 and Type 1 transitions outlined and a definition of the three zones: south, central and north. The pink lines indicate Type-0 transitions between neighboring FADs within the same zone, and the yellow lines indicate Type-1 transitions between neighboring FADs in different zones.

**Table S1**: The experimental occurrence, number of bonds in the array and the assigned probability for inter-FAD movements according to the different types.

| Type | Occurrence | # Bonds | Assigned Probability |
|---|---|---|---|
| 0 | 88 | 9 | 0.39 |
| 1 | 53 | 8 | 0.27 |
| 2 | 34 | 11 | 0.12 |
| 3 | 18 | 6 | 0.12 |
| 4 | 5 | 2 | 0.10 |